# Radiation conditions in relativistic interstellar flight

Oleg G. Semyonov

*STONY BROOK UNIVERSITY, STONY BROOK, NY 11794, NEW YORK (USA)*

Abstract. Radiation hazard on board of a relativistic rocket can be of internal and external origin. Because of its highest specific energy density, antimatter is considered to be the preferred rocket fuel for accelerating a multi-ton rocket up to relativistic speed. High-energy products of matter-antimatter annihilation ($\gamma$-photons and meson radiation) can produce a severe radiation hazard for crew and electronics. Two factors can stand against our pursuit to the stars: 1) cooling of a multi-GW propulsion engine, which can be done in space by thermal radiation only, and 2) intense nucleonic radiation originated from the oncoming relativistic "headwind" of interstellar gas as well from cosmic rays which intensity increases with rocket velocity. When a rocket accelerates to a relativistic speed, the rarefied interstellar gas of neutral and ionized atoms and molecules transforms into the oncoming flux of high-energy nucleons irradiating the rocket body and creating the severe radiation hazard. The oncoming flux of relativistic dust granules imposes a threat of mechanical damage to the rocket body. Possible protection measures are discussed.

## 1 Introduction

Technical and physical problems inherent in relativistic interstellar flight with an energy source and propellant on board of a starship are considered in details in [1]. Here we discuss one physical factor we will inevitably meet on our road to the stars: intense ionizing radiation originated from a) propulsion engine, b) relativistic "headwind" of high energy electrons, nuclei, atoms, and molecules of interstellar gas and c) galactic cosmic rays. It is well known that chemical, magnetohydrodynamic (MHD), and nuclear rocket engines are unable accelerating a multi-ton rocket to relativistic speed above 0.1c, where c is the speed of light, because of their relatively low energy capacity. In general, the higher is the specific energy density of a fuel, the lesser fuel reserve and therefore smaller launching mass of a rocket is needed to accelerate a rocket to the desired speed with the same mass rate of fuel consumption. Propulsion exhaust velocity $v_j$ also matters: the conventional rocket engines produce a copious mass exhaust with relatively small exhaust velocity, which results in fast fuel and propellant consumption in accordance with the expression for the mechanical thrust $F = v_j(dM/dt)$, where dM/dt is the exhaust mass flow per unit time. The achievable speed of chemical and MHD rockets is in the range from several to tens kilometers per second. The higher is the exhaust velocity $v_j$ , the lower is propellant consumption rate to get the same thrust and the higher speed of the rocket can be achieved. The rockets powered by nuclear reactors and propelled by high-energy ions (high-velocity plasma) effux [1, 2] can optimally reach a speed of several hundred km/s in few months. To accelerate a multi-ton rocket to a relativistic speed, the option is to produce a sufficient thrust for tens of years of flight using a propulsion engine that generates a relativistic exhaust jet powered by antimatter annihilation. When propellant velocity $v_j$ becomes relativistic, the expression for the thrust $F$ transforms into $F = \gamma_j\beta_j c(dM/dt)$, where $\beta_j = v_j/c$, $\gamma_j = (1 - \beta_j^2)^{-1/2}$, and $c$ is the speed of light in vacuum. Antimatter is a fuel of highest specific energy density because virtually all mass of annihilating matter and antimatter can be converted into energy [1 – 4]. Two concepts of annihilation-powered relativistic rocket have been suggested: a) propulsion by the products of matter-antimatter annihilation in a kind of nozzle (direct annihilation propulsion) [3, 4] and b) propulsion by high-energy ions accelerated in a thruster powered by annihilation reactor (relativistic ion propulsion) [1, 2]. Two direct annihilation propulsion thrusters have been commonly discussed: photon rocket propelled by a beam of γ-photons emitted in the process of electrons-positrons annihilation [3] and meson rocket propelled by a flux of π-mesons or μ-mesons produced by annihilating protons and antiprotons [4]. Relativistic ion propulsion engine [1, 2] includes an annihilation reactor to generate electrical energy for powering a set of relativistic ion accelerators (thruster) which produce the exhaust beam of high-energy ions of conventional matter

Photon rocket carries a fuel containing positrons which are supposed to annihilate with electrons at the focal spot of a photon-reflecting mirror in order to produce the exhaust beam of γ-photons. Meson rocket carries an antimatter fuel annihilating with ordinary matter in a magnetic nozzle to produce an exhaust jet of charged π-mesons for thrust. Virtually all the products of matter-antimatter annihilation (γ-photons, mesons, electrons, or other high-energy particles) are hazardous for astronauts and electronics, if they leak from the annihilation zone and irradiate the rocket body. Possible exception is the flux of neutrinos freely escaping the propulsion engine because of their weak interaction with matter. At relativistic speed of the rocket, innocuous interstellar gas forms an oncoming relativistic flux of atoms and ions perceived as hard ionizing radiation which is highly dangerous for crew and electronics and requires a robust frontal shield to absorb or deflect the relativistic headwind of high-energy nucleons [1]. Cosmic rays and cosmic γ-radiation add to the radiation hazard and may also require protective measures. The grains of interstellar dust turn into relativistic micro-projectiles bombarding the frontal parts of a relativistic starship and causing mechanical damage [1]. Many technical problems must be solved before we risk flying with a relativistic speed beyond the solar system and among them the problem of shielding of a spacecraft from ionizing radiation of internal and external origin as well as from damaging bombardment by oncoming relativistic dust is among the most challenging.

## 2 Radiation from propulsion engine

Two possible techniques for realization of direct propulsion by the products of matter-antimatter annihilation has been considered commonly: 1) photon rocket propelled by γ-photons born in the process of electron-positron annihilation at the focal spot of a mirror [3] and 2) meson rocket propelled by a jet of charged π--mesons produced in the process of protons-antiprotons annihilation inside a magnetic nozzle [4].

*2.1 Photon rocket*

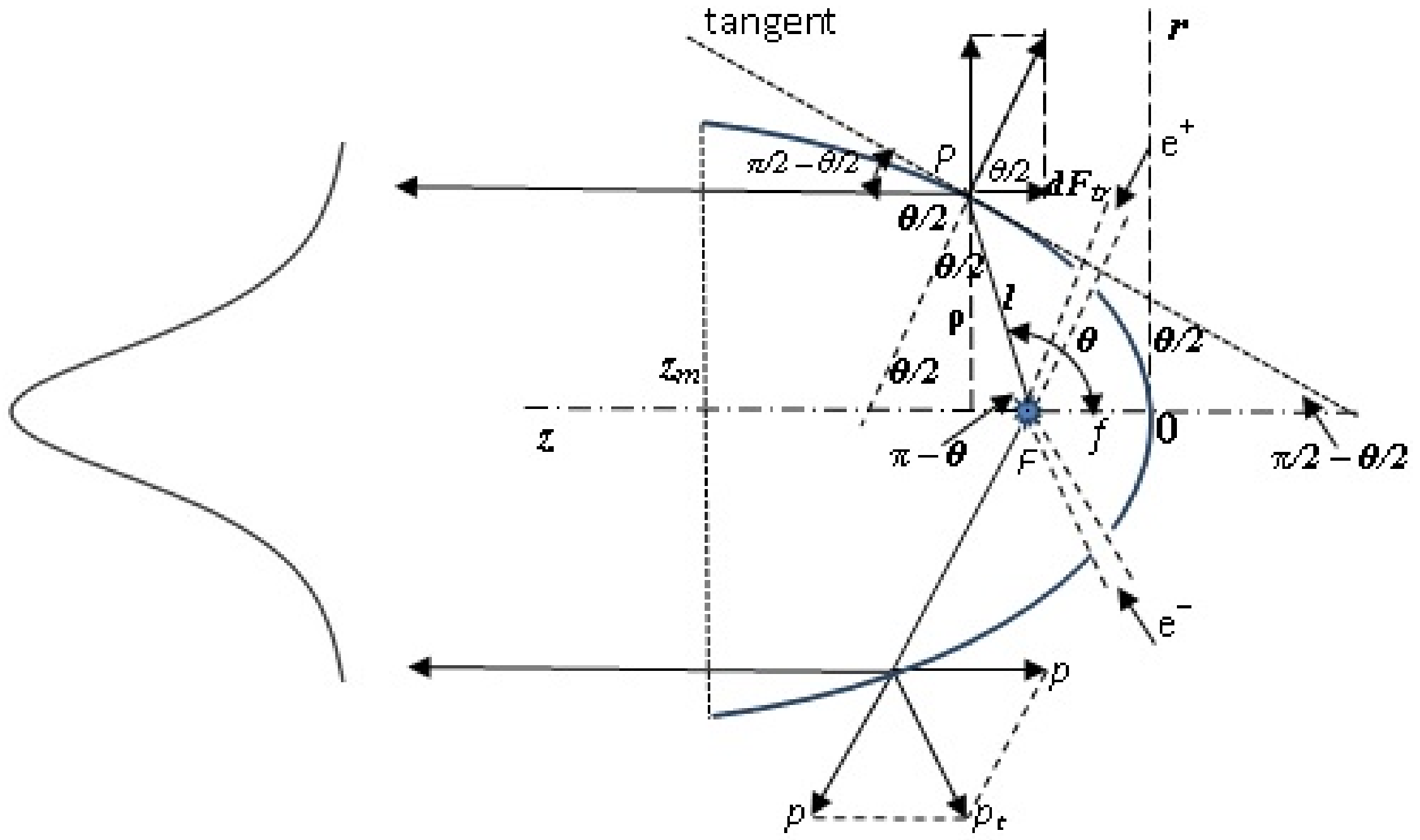

Fig.1 Schematic cross-section of photon-propulsion engine with a parabolic mirror (reproduced from [1]). The focused beams of electrons e⁻ and positrons e+ are inserted from the sides of a mirror to cross at the focal spot of a parabolic mirror. After electron-positron annihilation at the beam crossing region, the emitted γ-photons reflect from the mirror and form an exhaust beam. The radial distribution of energy density in the photon beam is shown to the left of the mirror. The details and discussion can be found in [1, pp. 14–15].

Two or several electron and positron (antielectron) beams cross in the focal spot of a photon-reflecting parabolic dish (Figure 1). Each act of electron-positron annihilation releases two γ-photons with their energy of the order of 0.5 MeV in opposite directions in the rocket coordinate frame, and one or both photons impact the parabolic mirror depending upon the axial extent of the dish so that each of the reflected photons transfers to the mirror (depending on the angle of incidence at a given point on the mirror surface) a portion of its mechanical momentum h$\nu$/c = h/λ, where h is the Plank constant, $\nu$ is the frequency of electromagnetic wave associated with each emitted photon and λ is the wavelength of this electromagnetic wave. Assuming all electrons and positrons annihilate at the spot of beams crossing with its size small in comparison to the focal distance and the overall size of the mirror, the emission spot can be treated as a point source of γ-photons creating an almost ideally parallel beam of photons after their reflection from the mirror.

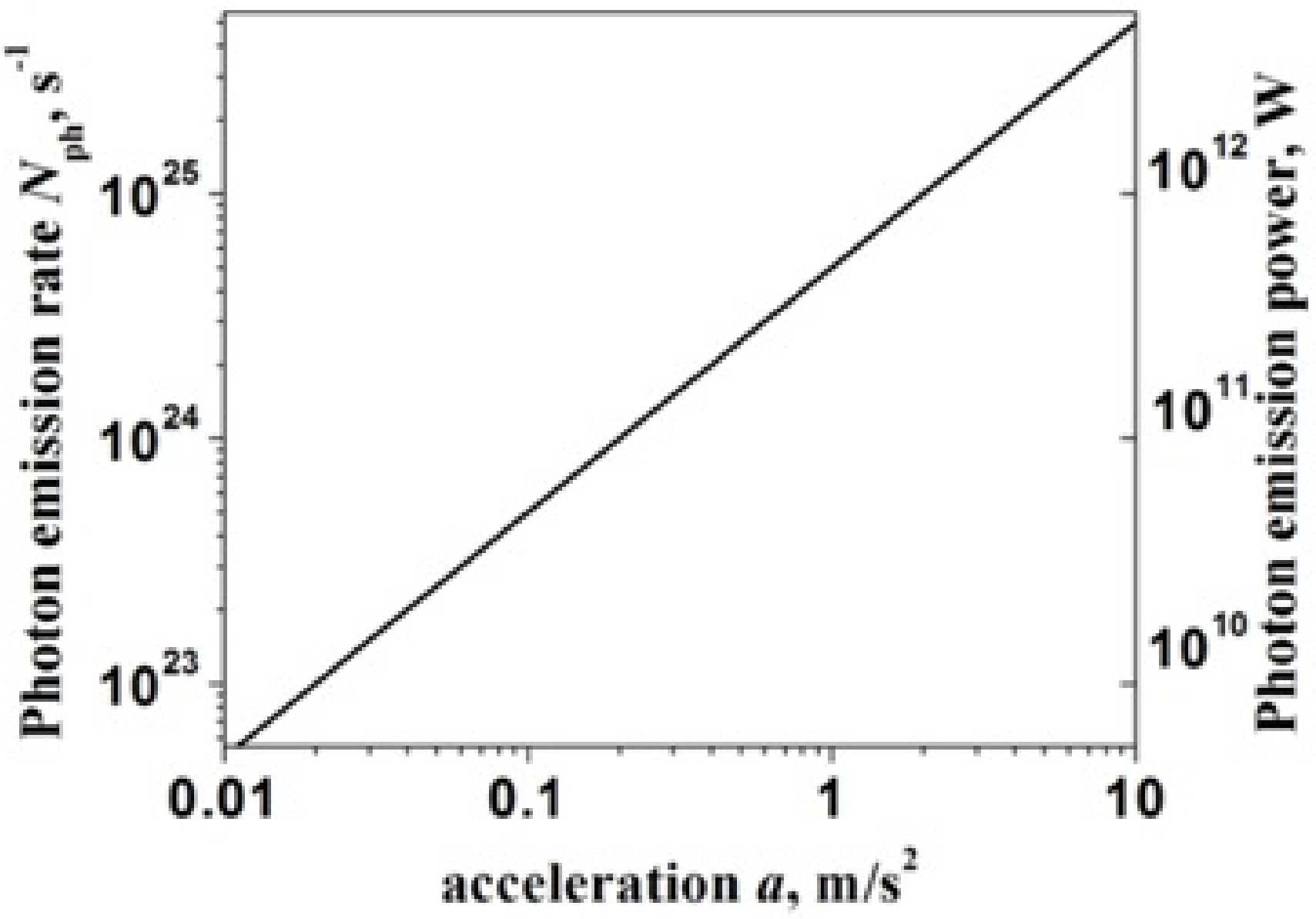

Fig.2 Photon emission rate $N$ per second and emission power (watts) per one ton of the photon rocket mass as a function of proper acceleration provided a hundred-percent parabolic mirror (reflection coefficient R = 1) is implemented for the efflux beam formation. The length of the mirror is taken equal to its focal distance (reproduced from [1, p. 18]).

Radiation hazard, photon emission rate, and power of the flux of γ-photons from the photon thruster can be estimated from the rocket equation for a chosen rocket launching mass and engine power [1, pp. 16–17].

$$\frac{M}{M_0} = \left(\frac{1-\beta}{1+\beta}\right)^{\frac{1}{\mu(1+R)\Phi}} \quad (1)$$

where $M_0$ is the launching mass, M is the instant mass, $\beta = \frac{v}{c}$ , $\mu = N_{ph}/N_{e+e-}$ is the annihilation efficiency in the spot of $e^+e^-$ beams crossing, R is the reflection coefficient of the mirror, $\Phi = 1 - \frac{1}{\left(1+\frac{z_m}{f}\right)^2}$, $z_m$ is the length of the mirror, and $f$ is its focal distance. Emission rate of photons and photon efflux power are shown in Figure 2 as functions of the rocket acceleration per one ton of the rocket mass. To produce acceleration of 1 $m/s^2$ (one tenth of free-fall acceleration on the Earth's surface), the propulsion power of a hundred-ton rocket must be of the order of 100 TW which corresponds to the 0.5-Mev photon emission rate of the order of $10^{27}$ (ten to the power of 27!) photons per second. Obviously, the combined flow rate of electron and positron beams into the annihilation spot must be of the same order of magnitude. The photon power flux is enormous and such a thruster is hardly imaginable at a hundred-ton spacecraft (for comparison, our industrial power plants supplying electricity to industrial facilities and big cities are in the range from gigawatts to tens of gigawatts).

This idealistic design with a hundred percent mirror is unrealizable for yet another reason. The wavelength of electromagnetic waves corresponding to 0.5 MeV photons is below the inter-atomic distances in all known materials thus no known material can respond to such high-frequency electromagnetic waves as a medium characterized by its refractive index and reflection coefficient which emerge for low-frequency light because of the collective reaction of many atoms and molecules to the relatively long electromagnetic waves. It means almost no reflection of 0.5 MeV photons from the known materials. Photon-absorbing dishes [3] are thinkable but the dish material must absorb, withhold, and dispose to space all the power of the photon flux otherwise either the rocket itself will be irradiated by an enormous flux of γ-radiation or a thick and heavy dish is required together with a huge thermal radiator to dispose the thermal energy to space (cooling in space vacuum can be done by thermal radiation only). It should be noted also that transportation and focusing of high-current electron (positron) beams in vacuum is not an easy task and their annihilation cross-section in realistic conditions is quite small to count on complete annihilation of antihydrogen in the focal spot of the mirror. Either annihilation will be incomplete or the size of annihilation zone must be of hundreds of meters or more thus no parallel photon exhaust beam can be formed unless a mirror is of many kilometers in size [1, 5]. A hope for positronium storage on board (quasi-atoms built by an electron and a positron) seems to be futile because no stable material containing relatively short-living positronium atoms has ever been suggested. The concept of photon rocket powered by electron-positron annihilation meets many unresolved problems and hardly realizable in practice.

### *2.2 Meson rocket*

Radiation hazard can arise from annihilation products escaping a magnetic nozzle to the rocket body [1, pp. 23-32]. The idea of magnetic nozzle stems from thermonuclear research on magnetic traps (magnetic bottles), in which high-temperature plasma of charged particles can be contained. Magnetic field in the magnetic nozzle is configured to form a jet of charged products of matter-antimatter annihilation [4]. According to the concept, two or several beams of protons and antiprotons cross inside a chamber, where a gradient magnetic field is induced by a system of current-carrying coils to produce mainly longitudinal magnetic field with its intensity B diminishing to the exhaust end of the chamber (Fig.3 and 4).

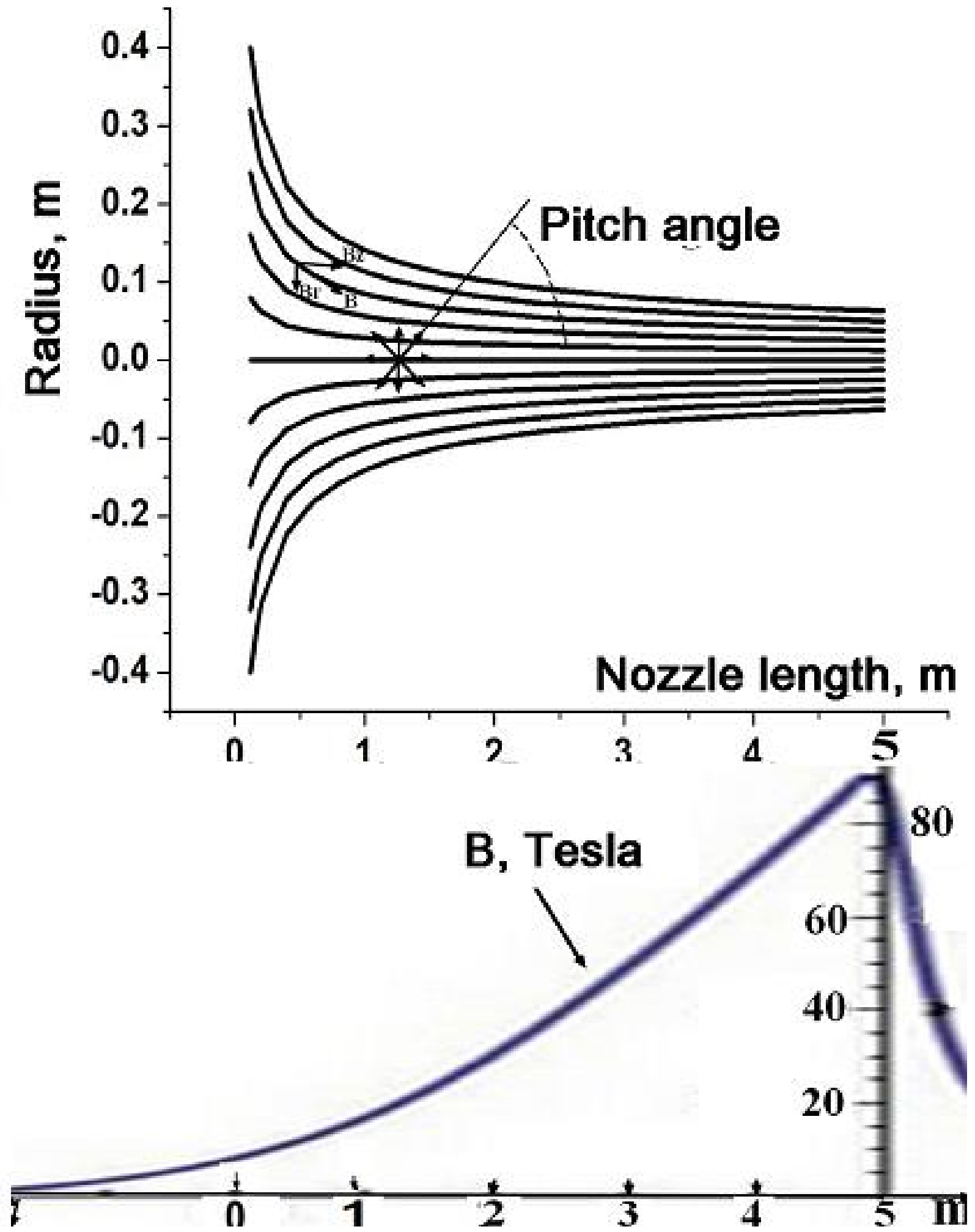

Fig.3. Magnetic lines in a five-meter nozzle (Figure 4 below) with the gradient magnetic field dB/dz = 19.98 T/m along the z-axis assuming B0 = 0.1 T at z = 0 (at the exhaust end). A point-like source of π-mesons is on the z-axis closer to the exhaust. The axes are measured in meters. Shown below is the distribution of magnetic field inductance B along the z-axis directed along the magnetic field gradient and the thrust vector. Adapted from the reference [1, figures 1.9 and 1.14]

Each proton-antiproton pair annihilates on average into five π-mesons with three charged π-mesons and two neutral π-mesons. Each neutral π-meson virtually instantly decays into two γ-photons with their energy about 200 MeV. Charged π-mesons (pions) decay (their decay time about 70 ns) into correspondingly charged μ-mesons (muons) and neutrinos. Positively charged muons decay into positrons and antineutrinos while every negatively charged muon decays into electron and neutrino. Mechanical momentum of charged annihilation products can be used to produce thrust provided a magnetic field of appropriate configuration is induced in the nozzle to force the charged particles to drift predominantly toward to the exhaust end of the nozzle and to form an efflux jet. In the configuration of predominantly longitudinal magnetic lines with a gradient magnetic inductance (Fig. 3), the charged products of proton-antiproton annihilation gyrate in the magnetic field and drift along the magnetic lines. The longitudinal "force" that acts on a gyrating particle is opposite to the magnetic field gradient –dB/dz. It slows down the particles drifting originally in the direction of a stronger magnetic field (initially emitted at a pitch angle below 90 degrees, i.e. to the right in Fig.3), and accelerates their drift, when they move down the slope of the magnetic field. A possible configuration of current-carrying coils to form a gradient magnetic field with the almost linear increase of magnetic inductance along z-axis to produce the thrust by a jet of π-mesons only is shown in Fig.4 [1]

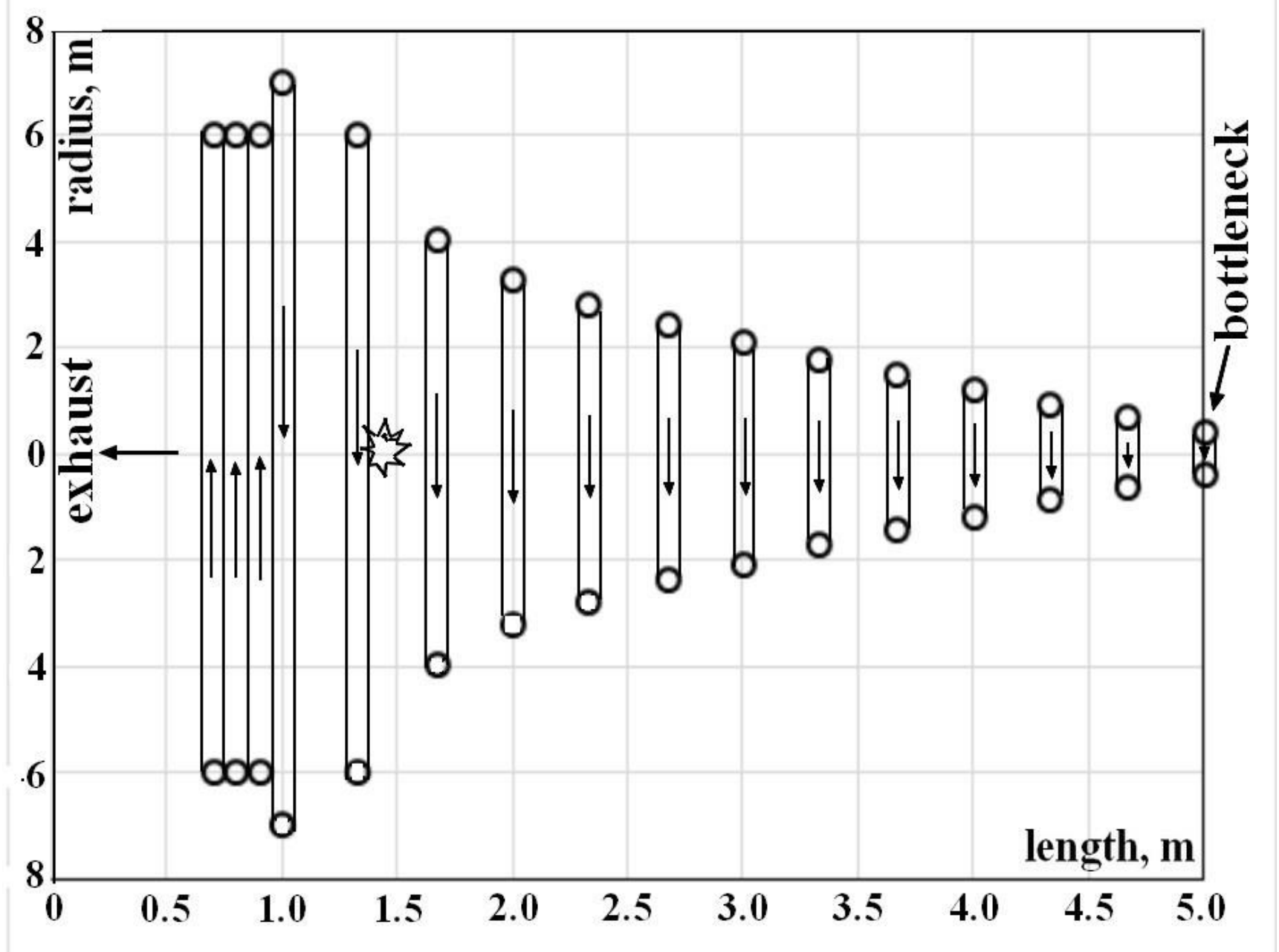

Fig.4 Five-meter long magnetic nozzle to produce thrust by π-mesons. It consists of current-carrying loops with their radii increasing to the exhaust end to produce an almost linear slop of magnetic inductance B along the z-axis toward the exhaust end z = 0 (Fig.3). The directions of current are shown by arrows. To dump the tail of magnetic field beyond the exhaust end as much as possible, three additional loops with the opposite direction of current are added to the exhaust end. The position of the source of pions for calculations of the pion trajectories (Fig.5) is marked by a star. Adapted from the reference [1, figure 1.14]

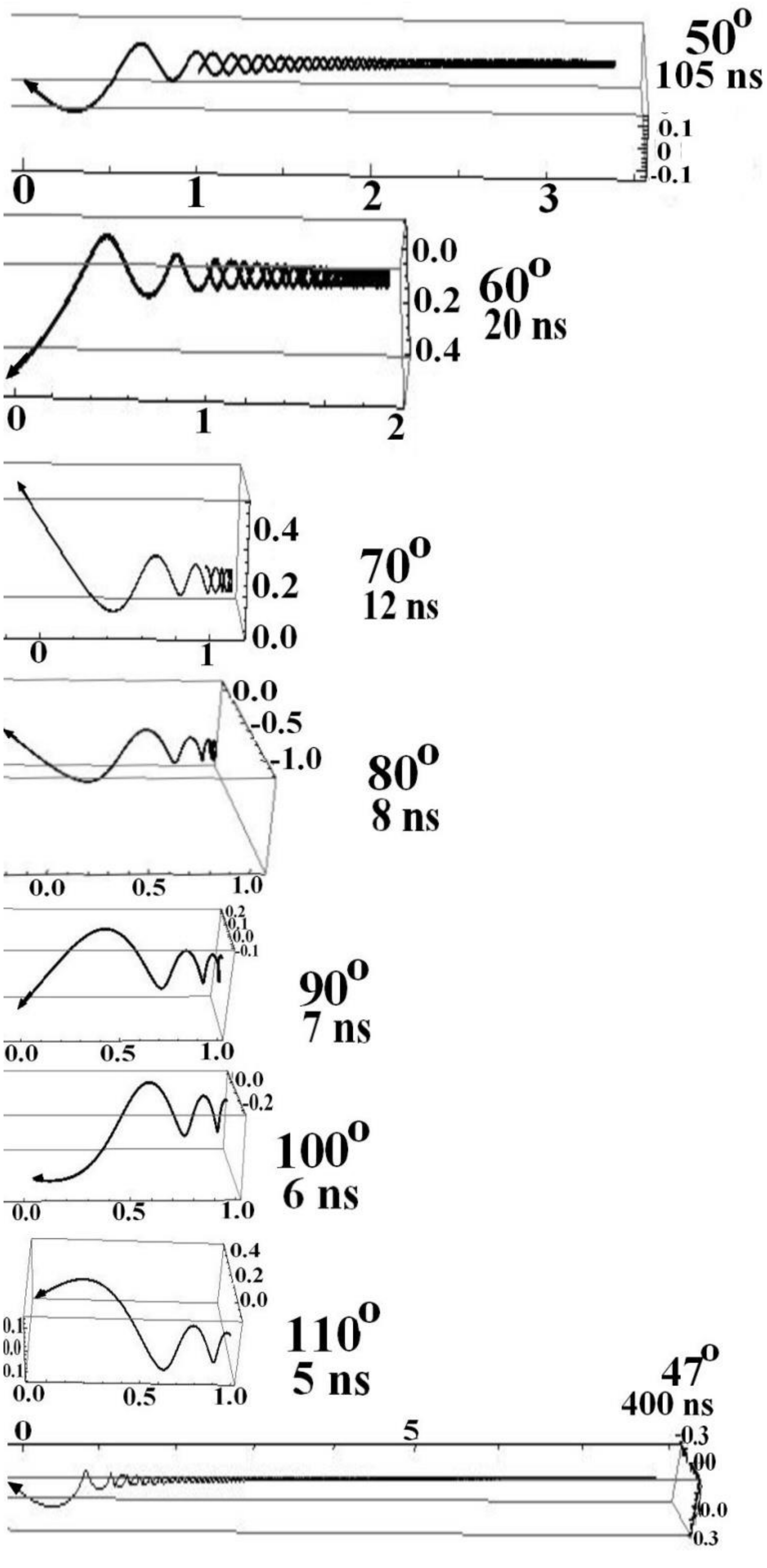

Fig.5 Trajectories of a positively charged π-meson in a five-meter magnetic nozzle with a linear gradient of magnetic field ($B_m$ = 100 T) calculated for different initial pitch angles $\theta$ (angle of emission of π-mesons relative to z-axis as shown in Fig.3). The z-axis is directed along the magnetic field gradient. Time of flight of π-mesons from the source (annihilation zone) positioned at the distance of one meter from the exhaust end of the nozzle is shown to the right of each trajectory for each initial pitch angle of emission $\theta$. Adapted from [1]

Trajectories of π-mesons emitted from a point on z-axis of this magnetic five-meter nozzle are shown in Fig.5 for different initial pitch angles. Calculations were performed assuming the linear gradient dB/dz from B = 0.1 T at z = 0 (z = 0.5 m in Fig. 4) to B = 100 T at z = 5 m (bottleneck). The average time of flight of charged pions along the trajectory (before 50% of them decay into μ-mesons) is about 70 ns in the rocket coordinate frame so the nozzle's length of 5 meters is chosen for simulations from the estimate of the full travel $s_\pi$ of pions along their spiral trajectories about 20 meters before their half-decay. To produce the thrust by π-mesons predominantly, the length of magnetic nozzle between the point of maximum magnetic field and the exhaust end should not exceed five meters in order to give them a sufficient time to exit the nozzle before their decay, if emitted originally at the relatively small pitch angles. A better solution can be the significantly longer nozzle, in which almost all charged π-mesons decay into μ-mesons which have a relatively longer decay time. This way, virtually all charged π-mesons drifting to the exhaust end can produce thrust and yet an additional thrust can be produced by μ-mesons gyrating in the magnetic field [1, pp. 38–41]. The length of this magnetic nozzle can be of tens and even hundreds of meters.

Magnetic mirrors have their own essential inherent drawbacks. Firstly, it is impossible to inject the beams of protons and antiprotons from the sides of a magnetic nozzle because the charged particles cannot propagate across the magnetic lines. The only possibility is to inject the beams through the nozzle's bottleneck (the nozzle's edge with the maximum magnetic field) along the z-axis, i.e. from the right in Fig.3 and 4. Both beams must follow the same way and be pretty thin (small diameter) otherwise they will be redirected by the strong radial component of the magnetic field in front of the entrance into the magnetic nozzle and never cross to annihilate in the magnetic nozzle [1, p. 41]. Secondly, the calculated π-meson trajectories [1] demonstrate practical impossibility to create a nearly parallel exhaust jet of π and μ-mesons with their exhaust velocity vectors closely aligned along the z-axis. It means the reduced thrust and lower rocket acceleration in comparison with an ideally aligned exhaust jet of annihilation products. Another essential drawback of the magnetic mirror is the inevitable leak of charged mesons initially emitted at the small pitch angles through the bottleneck. Even if we manage to tightly focus the beams and produce an ambiplasma (mixture of both beams) in which all protons and antiprotons annihilate inside the magnetic nozzle, the mesons with their original pitch angle $\sin \theta < (B/B_m)^{1/2}$, where B is the magnetic inductance at the point of their emission and $B_m$ is the maximum magnetic inductance at the bottleneck, will never be reflected to the exhaust end but continue to travel to the rocket body producing firstly a braking thrust and secondly creating a severe radiation hazard for crew and electronics in addition to γ-radiation emitted by the decaying neutral π-mesons [1, p. 36]. Positioning the point of proton-antiproton annihilation (annihilation zone) closer to the exhaust end of the magnetic nozzle will reduce the thrust produced by the π-mesons emitted at the initial pitch angles $\theta > 90$ degrees. Shifting it closer to the bottleneck will enlarge the loss-cone of π-and μ-mesons escaping through the bottleneck thus reduce the thrust, too. The only foreseen protection against the flux of γ-radiation is a shield of γ-absorbing material.

The relativistic rocket equation for direct propulsion by mesons ([1], p. 19) is

$$\frac{M}{M_0} = \left(\frac{1-\beta}{1+\beta}\right)^{\frac{1}{2\beta_{ex}}} \tag{2}$$

where $M_0$ is the launching mass, $M$ is the instant rocket mass is the rocket map-velocity in the reference coordinate frame, $\beta = v/c$, and $\gamma = (1 - \beta^2)$, $\beta_{ex} = u_{ex}/c$, and $u_{ex} = \mu\gamma_\pi\beta_\pi c$ is the effective exhaust speed in the rocket proper coordinate frame. The effective exhaust speed $u_{ex}$ is lower than the average speed of charged pions $v_\pi = 0.93\ c$ because only a fraction $\mu\gamma_\pi \leq 0.622$ of the initial proton+antiproton mass converts into the pions total mass-energy. To get the acceleration $a = 1$ m/s$^2$ of a thousand-ton rocket, the total annihilation power should be of the order of $2\times10^8$ MW and the kinetic power of the meson efflux about $5\times10^7$ MW [1, p. 56]. The emission rate of 200-MeV γ-photons from decaying neutral π-mesons will be of the order of $10^{24}$ photons per second, which corresponds to the radiation power of $6\times10^7$ MW. In order to reduce this huge flux of γ-photons to a relatively safety level, the rocket protecting shield of lead must be well above one meter in thickness. Such a shield would take a lion's share of the rocket dry mass. Positioning the propulsion engine sufficiently far from the control equipment and crew quarters can reduce the mass of the shield due to geometric reduction factor but the rocket axial elongation to tens kilometers or more will be hardly acceptable. According to the calculations [1], the loss-cone of π-mesons (pions) through the bottleneck of the nozzle with the maximum magnetic field of 100 Tesla is about or wider than 1 steradian thus ten or more percents of π- and μ-mesons will leak through the bottleneck to the rocket body creating a huge radiation hazard in addition to γ-radiation. To screen the rocket from the flux of charged mesons, a magnetic shield analogous to the protective shield against the oncoming nucleonic radiation originated from the flux of relativistic interstellar gas (see Section 3 below) can be mounted between the nozzle and the rocket body to absorb or deflect the charged mesons. It will not however eliminate the need for a shield made of dense and heavy material against γ-radiation. Taking into account the problem of injection of proton and antiproton beams into the magnetic nozzle and a huge practical length of annihilation zone (tens or hundreds of meters) for the achievable diameter of proton and antiproton high-current beams [1], the direct propulsion by the annihilation products seems to be not a promising solution for interstellar relativistic rockets.

*2.3 Relativistic ion propulsion*

Alternative antimatter-powered propulsion has been discussed in [1, 2]. According to the conception, an antimatter-annihilation reactor can be used for electrical energy production to power a high-energy ion thruster. Basically, any energy-generating reactor (nuclear, thermonuclear, or antimatter annihilation reactors) can be utilized for thrust production by a jet of plasma or accelerated ions. Because of its highest energy density per unit mass of annihilating matter and antimatter (fuel) thus much lower rate of fuel consumption to generate the same power, antimatter reactor seems to be indispensable for relativistic interstellar spacecrafts. It supplies power to one or several ion accelerators of conventional matter which produce the efflux beam of high-energy ions. From the expression for momentum conservation, the thrust $F = N\gamma_i m_0 \beta_i c$, where $N$ is the exhaust rate of ions per unit time in the rocket coordinate frame, $m_0$ is the mass of rest of the ions in the exhaust jet, $\beta_i = v_i/c$, $v_i$ is the exhaust speed of ions in the rocket coordinate frame, and $\gamma_i = (1 - \beta_i^2)^{-1/2}$, the thrust increases with the growing exhaust velocity through the factors $\gamma_i$ and $\beta_i$. allowing the exhaust rate $N$ reduction thus propellant economy. The price is that the efflux kinetic power $W = Nm_0c^2(\gamma_i - 1)$ also grows with the efflux velocity through the factor $\gamma_i$ and tends to be comparable with the mass-energy $m_0c^2$ or even exceeding it, if $\gamma_i > 2$. The ratio $F/W$ is a diminishing function with the increasing $y_i$ [1, 2] therefore a higher reactor power is needed to obtain the same thrust with the increased exhaust speed of ions. The reactor power (less losses) goes to kinetic energy of almost completely aligned relativistic beam of ions. To compensate the positive charge of the ion exhaust beam, emitters of electrons is to be mounted around the exhaust end of the ion accelerators in analogy with the ion thrusters already in use at interplanetary probes.

From the rocket equation [1, 2],

$$\frac{M}{M_0} = \left(\frac{1-\beta}{1+\beta}\right)^{\frac{(1+(\chi_b-1)/\varepsilon)}{2\gamma_b\beta_b}} \tag{3}$$

where $M_0$ is the rocket launching mass, $M$ is its instant mass, $\beta = v/c$, $\gamma_b$ and $\beta_b$ are the relativistic Einstein factors of the efflux ions in the rocket coordinate frame, and $\varepsilon$ is a portion of fuel mass-energy that goes to the kinetic energy of the exhaust jet of ions, the achievable speeds $\beta_{0.5} = v_{0.5}/c$ and $\beta_{0.25} = v_{0.25}/c$ of a rocket at the moments, when either half of the rocket launching mass $M_0$ or three quarters of $M_0$ is exhausted for propulsion including the exhausted propellant and matter-antimatter burned in the reactor, are shown in Fig.6 as functions of the velocity factor $\beta_i = v_i/c$ in the exhaust ions for several propulsion efficiency coefficients $\varepsilon$ (efficiency of the annihilation reactor together with the gas turbines for electrical power production plus efficiency of the ion thrusters) [6]. The graphs are valid for any launching mass and propulsion power, however it should be remembered that the rocket acceleration $a$ and the time of flight $\tau$ in the rocket coordinate frame to the moments, when a half of the rocket launching mass (or three quarters of the rocket launching mass) is exhausted, depends on the rocket launching mass and the propulsion power. For illustration, the graphs of the rocket speed $\beta$ and the distance of flight $s$ during the acceleration period are given in [1, pp. 65–68] and [2] as functions of the time of flight measured by the rocket clock for the launching masses of 1000, 5000, and 10000 tons accounting for the rocket propulsion power of 1 TW and 100 TW.

The obvious advantage of relativistic ion propulsion powered by a reactor is that it gives much better freedom and flexibility in choosing the propulsion power and the propellant. In addition, it opens a possibility of independent control of kinetic energy of the exhaust ions and their mass flow. Any liquidized gas from hydrogen to xenon can be used for ion propulsion and these elements can be found almost everywhere in the universe. The increased exhaust velocity results in reduction of the propellant exhaust rate to get the same rocket acceleration. It allows achieving higher cruising velocity due to longer thrust with the same propellant reserve and even saving some propellant for braking and may be for the return flight. The price we have to pay for the increased exhaust velocity is either higher energy consumption to get the same thrust or lesser thrust and rocket acceleration with the same propulsion power thus longer time for picking-up the desired speed. Nonetheless, the possibility of achieving a higher rocket velocity at the moment, when a predetermined portion of rocket launching mass is exhausted, say, half of the rocket launching mass (Fig. 6), is beneficial because the total time of flight to a remote destination, including the period of cruising with a higher constant speed, can eventually be shorter.

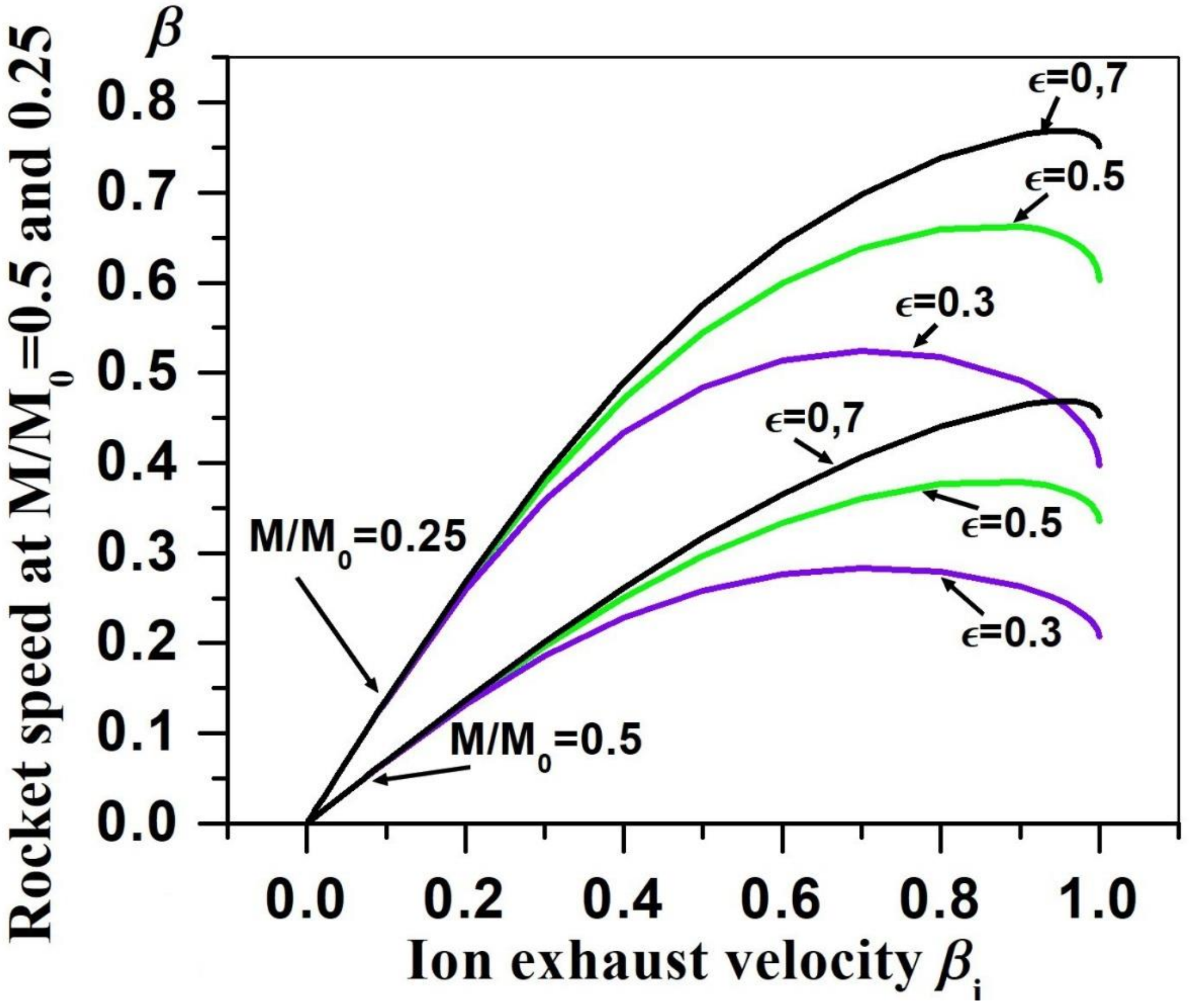

Fig.6. Map-velocity $\beta_{0.5} = v_{0.5}/c$ and $\beta_{0.25} = v_{0.25}/c$ of a rocket at the moments, when the residual mass of the rocket $M = 0.5M_0$ and $0.25M_0$, as functions of the proper velocity $\beta_i$ of the efflux of protons. The graphs are shown for the propulsion efficiency ε = 0.3, 0.5, and 0.7 and valid for any efflux power and launching mass. Adapted from the reference [6, figure 2]

A significant portion of mass-energy of annihilating atoms and antiatoms in the reactor can be converted to electricity. The inevitable loss is the flux of neutrinos and antineutrinos escaping freely to space (14.56% of the total mass-energy of annihilating protons and antiprotons). Another possible loss (additional 26% of annihilation energy) can be the flux of γ-photons emitted by the decaying neutral π-mesons. Unless the reactor's blanket effectively absorbs this γ-radiation or a shield is placed between the reactor and the rocket parts sensitive to γ-radiation, this gamma radiation can produce a severe radiation hazard onboard. Possible solution of γ-radiation problem is an annihilation reactor which contains a chunk of heavy-nuclei material with a high melting point (tungsten or natural uranium) irradiated by a stream of antiprotons or antihydrogen atoms (molecules). When an antiproton strikes a heavy nucleus, it annihilates with a proton or a neutron of the nucleus mostly near the surface of this nucleus and one or both γ-photons emitted by each decaying neutral π-meson can be absorbed inside the nucleus, if emitted in the forward-directed hemisphere [7]. The same is true for the mostly forward-directed charged π-mesons. The absorbed γ-photons and pions transfer their energy to kinetic energy of the nuclei fragments because γ-photons and pions entering a nucleus cause its excitation and probable fragmentation. The fragments lose their kinetic energy in collisions with electrons and nuclei in the material and heat it The charged pions emitted backward from the heavy-nuclei material should be absorbed in the blanket to utilize their kinetic energy, too.. Placing this heavy-nuclei module (prime heater) so that it is irradiated by antiprotons (antiatoms) from the aft of the rocket, it can simultaneously serve a shield to protect the rocket's parts positioned behind it (i.e., closer to the rocket nose) from mesons and gamma radiation. The reactor blanket and heavy-nuclei prime heater must be thick enough to absorb the high-energy mesons and gammas at least in the direction of the rocket parts that need protection. To compensate the loss of its material at the antimatter irradiated surface, the prime annihilation module can be designed porous or containing thin channels in it to deliver a liquid heavy-nuclei material with a lower melting point (melted metal or salt) to its surface and create a liquid layer on the irradiated surface, which actually annihilates with the incident antiprotons (antiatoms). This liquid material can also serve a primary cooler permanently replenished for its loss in annihilation at the irradiated surface. The prime heater with the heavy-nuclei prime cooler delivers the thermal energy released in annihilation to a heat exchanger, where a sort of gas is heated to be directed to the gas turbines for electrical power production. Neutrons generated in the process of heavy nuclei fragmentation can also add to radiation hazard onboard, if not absorbed in the material. [7]. Annihilation of antiprotons with heavy-nuclei may also result in many other physical effects not properly studied yet [8].

The launching mass of antimatter-driven rocket may be of hundreds to many thousands of tons including antimatter fuel and propellant stored in the corresponding tanks, protection means against hard radiation originated from the propulsion engine (annihilation nozzle or reactor) as well as produced by the relativistic 'headwind' of interstellar gas (see the section 3.1 below), power generating equipment, antimatter and propellant delivery systems to the reactor and/or propulsion engine, a means for exhaust jet formation, a heat-radiating cooler to dispose the heat of the engine to space, and many other mechanisms and auxiliary equipment. To accelerate a multi-ton rocket to a relativistic speed above $0{,}3c$ in a reasonable time interval of years or tens of years would require a multi-gigawatt or even multi-terawatt propulsion engine. [1] Irrespective of propulsion design, production and handling of such a huge power will be on the verge of technical possibilities even for the future advanced technology.

## : 3 Hard ionizing radiation of external origin

*3.1 Interstellar gas*

Outer space beyond the atmosphere is not empty void: it contains rarefied gas and dust. Interstellar gas is the necessary component of every galaxy and plays an important role in evolution of stars and galaxies. Permanently replenished by stellar wind (gas and plasma emanated from the stars in analogy to the solar wind) and as a result of catastrophic star explosions such as novas and supernovas, the clouds of interstellar gas give birth to the new generations of stars together with their planetary systems. After living through their life cycle, the stars replenish back the interstellar gas clouds to give birth to the next generations of stars (stellar recycling) [9]. Every galaxy is an evolving system of interdependent stellar objects and interstellar gas that fills each galaxy unevenly: there are relatively low-density regions and denser clouds. Our Sun was formed most likely in a dense gaseous cloud about five billions of years ago. Luckily for us, it is positioned now in a local low-density cavity about 400 light-years in size within the Orion spur of our Milky Way [10]. Interstellar gas contains 89% of hydrogen with 10% admixture of helium and about 1% of heavier elements such as carbon, oxygen, silicon, iron, etc. mostly accreted in dust granules [10, 11]. Concentration $n$ of neutral and ionized atoms and molecules in the vicinity of the Sun is about $3\times10^5$ m$^{-3}$.

When a rocket accelerates to a relativistic velocity v, all gaseous components and dust grains form a frontal flow of atomic and dust particles incident with a relativistic velocity on the rocket frontal parts (relativistic 'headwind') irrespective of the method of starship propulsion, its size, its geometry, and mass. This headwind of otherwise innocuous interstellar gas turns into an ongoing stream of high-energy nucleons and atoms while the dust granules become relativistic micro-projectiles bombarding the rocket hull. Kinetic energy of every particle relative to the rocket is $mc^2(\gamma-1)$, where m is the mass of rest (either a nucleon, gas atom, or dust grain), $\gamma = (1-\beta^2)^{-1/2}$, and $\beta = v/c$. Kinetic energy of ionized and neutral atoms of hydrogen – the main component of interstellar gas – exceeds 100 MeV at the rocket speed v > 0.5c, which is characteristic of high-energy nucleonic radiation analogous to the ion beams produced at high-power accelerator facilities. Despite extremely deep vacuum in interstellar space, the flux $P = \gamma n v$ of relativistic ions, atoms, and molecules relative to the rocket in the local cavity will exceed $10^9$ per square centimeter per second ($10^{13}$ per square meter per second) at the rocket velocity above $0.3c$. The rate of radiation dose absorbed in the tissue of an unprotected (unshielded) astronaut will exceed $10^4$ rems per second [1, 12] so the lethal dose of 1000 rems will be accumulated in his body in a fraction of second. The relativistic factor $\gamma$ in the expression for $P$ is due to the effect of relativistic time contraction. The flux of atomic particles and the dose rate for an astronaut without radiation protection are plotted in Fig.7 as functions of the rocket velocity factor $\beta = v/c$.

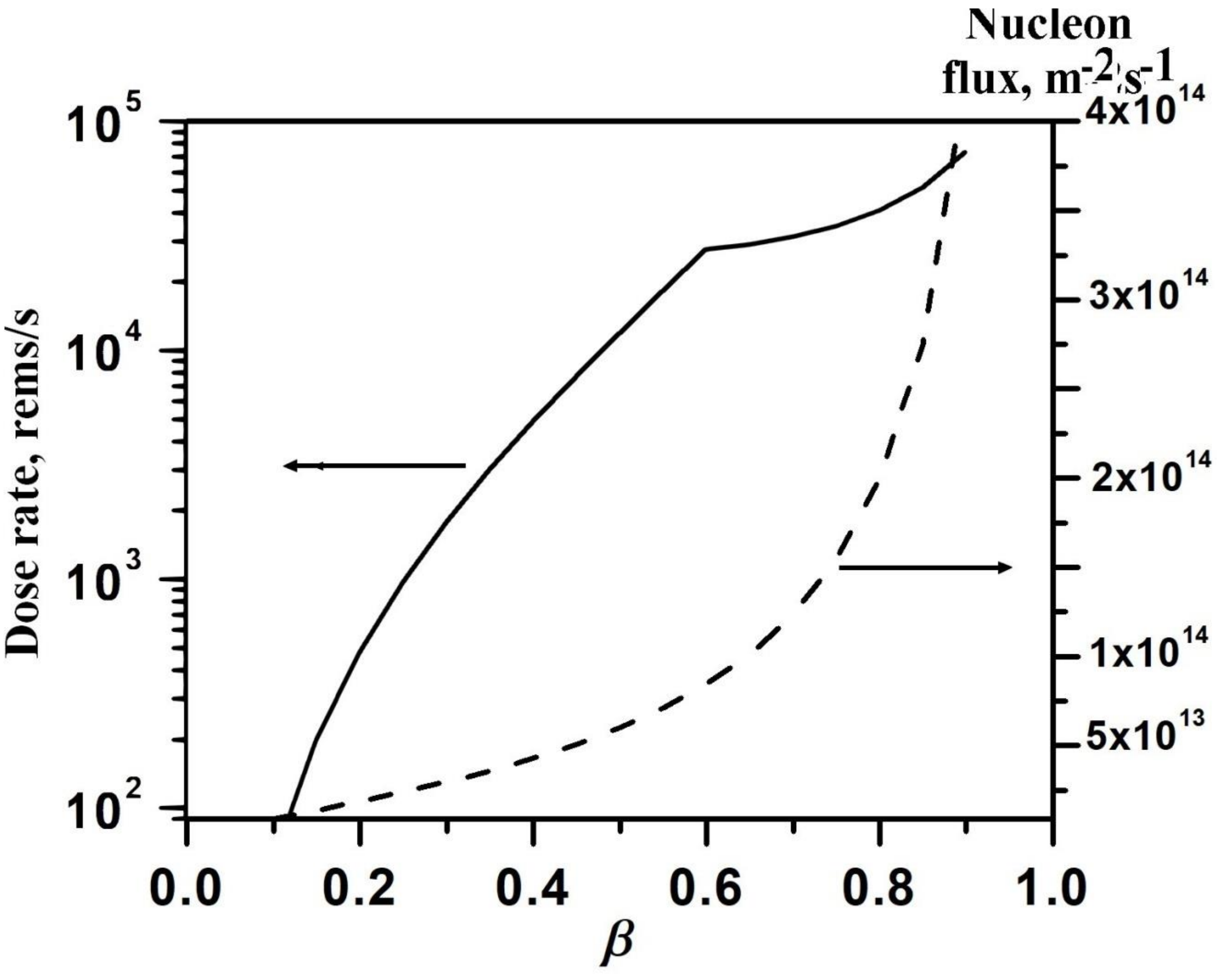

Fig.7. Flux of interstellar atoms and ions per square meter per second (dashed) incident on a rocket and the radiation dose rate (rems per second) obtained by an unprotected astronaut as functions of the rocket map-velocity $\beta = v/c$. A brake on the upper graph of the dose rate near $\beta = 0.6$ corresponds to the rocket velocity at which the penetration depth of the nucleons (protons mostly) in the tissue starts exceeding the average thickness of a human torso (about 30 cm). Adapted from the reference [6, figure 3]

Safe radiation dose is 5 rems according to the NIST safety regulations. The dose of hundred rems is considered dangerous due to high probability to develop cancer, and the dose of thousand rems or more is almost hundred percents lethal. To reduce the dose rate, a robust radiation-absorbing shield has to be mounted in front of the rocket. Material protective shield would require tens centimeters of iron or several meters of water or ice [1, p. 101], which means many additional tons to the rocket dry mass. The magnetic shield alone will not work because of a significant percentage of neutral components in interstellar gas. A relatively light-weight shield comprising a magnetic system and a thin electron stripper [1, 12] can protect the rocket from the relativistic flux of ionized and neutral components of interstellar gas. The electron stripper is a relatively thin solid disk installed at some distance in front of the rocket, which is mostly transparent for the high-energy nucleons but has a sufficient thickness to strip the electrons from the neutral atoms of incoming relativistic headwind (electron stripper). Magnetic system behind the electron stripper includes a solenoid that induces the magnetic field with the magnetic lines perpendicular to the rocket velocity vector by a winding made of superconductive wires [1, p. 112]. The coils can be made of thigh-temperature superconducting ceramics wound around a tank filled with a cryogenic liquid to form either toroidal solenoid producing the azimuthal magnetic field or a flat solenoid to generate the field with the strait magnetic lines. High-temperature superconducting ceramics is known to withhold the current density above 1 MA/cm2 and generate the magnetic field up to 30 T [5]. Combination of both geometries can protect the rocket body including the radiation cooler for waste heat disposal [1, pp. 110–112]. The flux of charged nucleons submerges into a tank filled with liquid hydrogen or helium through the relatively thin superconducting winding around the tank. The charged nucleons gyrate across the magnetic lines inside the tank and lose their kinetic energy in collisions with the atomic electrons and nuclei of the cryogenic liquid.

At a rocket speed below $0.8c$ and a magnetic field inductance of 10 Tesla, the radius of gyration of incoming H and He nucleons in the tank will not exceed one meter therefore the magnetic shield of two meters in thickness (significantly smaller than the full travel of nucleons along their trajectories in liquid hydrogen $s \sim 10$ m) will be sufficient for rocket protection. Possible accumulation of positive charge on the tank and on the rocket body can be compensated by accumulation of negatively-charged electrons on the electron stripper provided the magnetic system and the stripper are electrically connected. An additional advantage is that secondary μ-mesons and γ-radiation generated in the tank by the gyrating nucleons in their collisions with the nuclei of liquid that fills the tank will not be directed exclusively to the rocket body but distributed over $2\pi$ angle thus reducing their portion directed to the rocket and facilitating its protection against the secondary radiation. The sketch of conceptual relativistic ion-propulsion rocket containing the most important elements is shown in Figure 8.

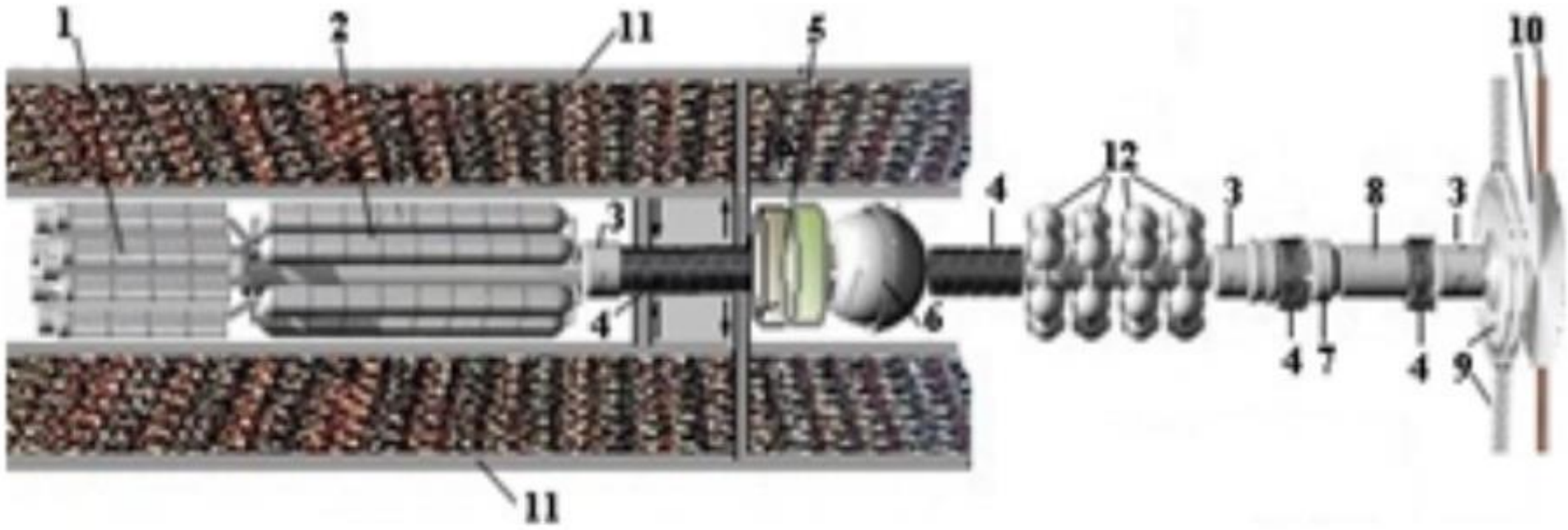

Fig.8. Conceptual relativistic ion propulsion interstellar rocket: 1 – ion thruster (an assembly of several ion accelerators producing the beams of relativistic ions); 2 – propellant tanks; 3 – low-temperature refrigerators; 4 – thermal insulation; 5 – gas turbines to generate electrical power; 6 – annihilation reactor; 7 – control bridge; 8 – crew quarters (if any) or auxiliary equipment; 9 – magnetic shields to protect the rocket body together with the thermal radiators from the ongoing headwind of charged nucleons; 10 – electron stripper of oncoming neutral atoms and absorber of oncoming free electrons in interstellar gas; 11 – thermal radiators for cooling the power-producing unit; 12 – antihydrogen tanks. Reproduced from [1, figure 3.7]

### *3.2 Cosmic rays and cosmic γ-rays*

Cosmic rays consist mostly of high-energy protons (90%) and α-particles (9%) bombarding an unmoving target uniformly from all directions [13]. Their energy maximum lies between 300 MeV and 1 GeV. The radiation hazard from cosmic rays is tangible both for non-relativistic and relativistic space flights. Strictly speaking, complete shielding against cosmic rays would require something analogous to Earth's atmosphere for example a shell of water of 5 m in thickness around the rocket [14]. This is not a welcomed solution both for interplanetary and interstellar flights because of the significant increase of the rocket dry mass. Even a water shell of 1 m in thickness, satisfying radiation safety standard, could be excessively heavy. In addition, a layer of dense material will be needed to absorb the penetrating secondary gamma and muon radiation generated in collisions of cosmic rays with nuclei of the shielding material. If the NASA's limit of 400 rems per individual during his duty (meaning doubled probability to develop cancer) will be accepted for interstellar flights, a thinner and therefore lower-mass material shield can be accepted for the short-term missions (1 to 5 years). The life-long interstellar travels of astronauts will definitely require almost complete shielding of crew quarters.

In analogy with the phenomenon of relativistic aberration of light from the point of view of a spacecraft moving with a relativistic speed due to transformation of the incident angles from the map-frame to the comoving coordinate frame [15, 16], the equation for transformation of the angles of incidence of relativistic massive particles isotropically moving in space in all directions can be derived [1, 12]. A frontal shield installed for protection of crew and electronics from relativistic headwind of interstellar gas can also absorb a portion of cosmic rays because of their increasing beaming when the rocket speed is close to the speed of light. However, this relativistic beaming effect is not as significant at the achievable rocket speed up to 0.7c to hope on a significant reduction in cosmic rays intensity from the sides. Accepting the average radiation quality factor Q = 6.5 for the cosmic rays (Q = 5 for protons and Q = 20 for α-particles according to European Nuclear Society[1]) the estimated annual equivalent radiation dose accumulated in the astronaut's body from unshielded cosmic rays D ≈ 30$N$ rem/year is plotted in Figure 9 as a function of the rocket velocity factor $\beta = v/c$, where $N$ is the flux of cosmic rays per square centimeter per second integrated over the angles of incidence [12] and [1, p. 106].

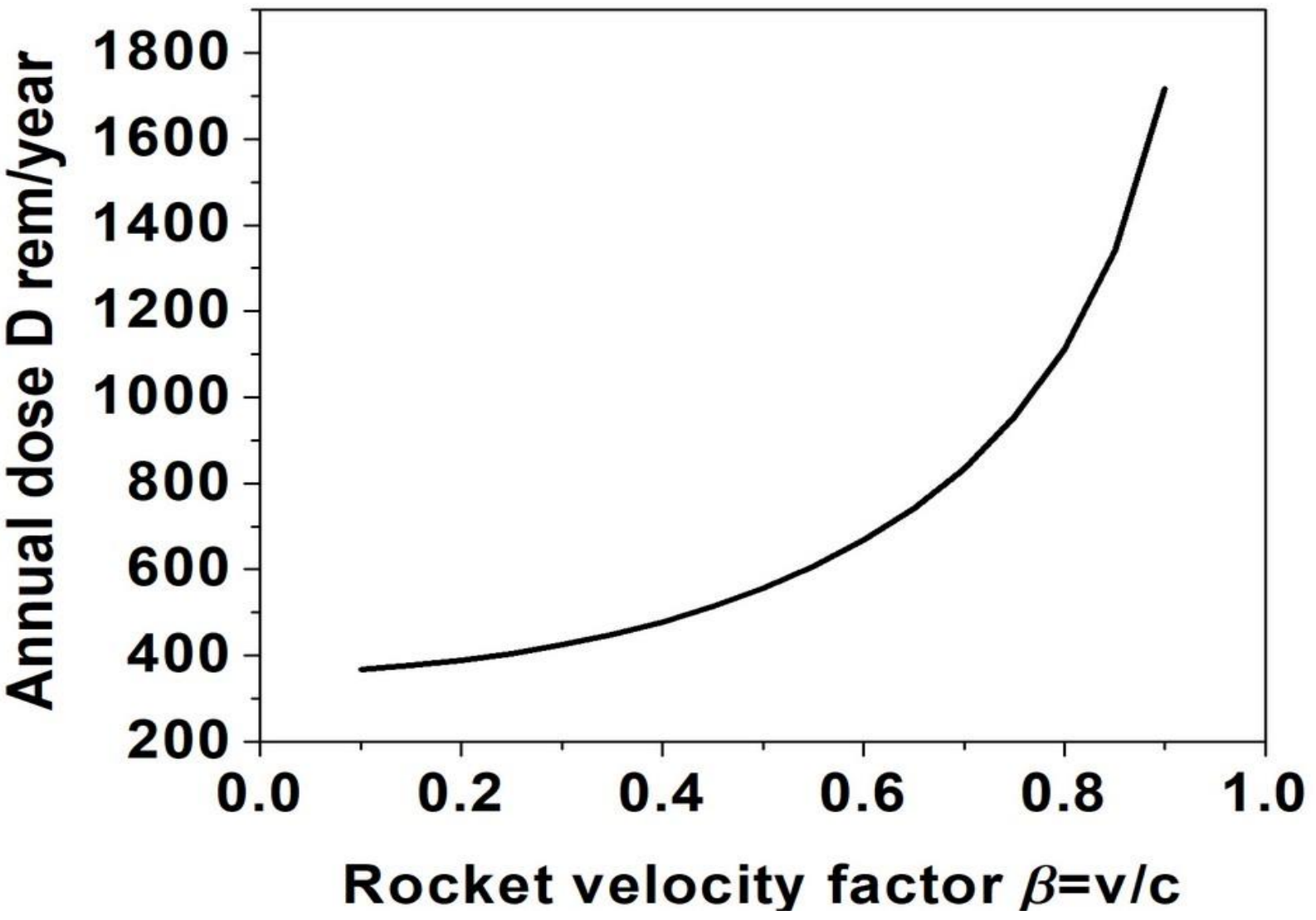

Fig.9. Annual dose accumulated in an unprotected astronaut body from cosmic rays as a function of the rocket velocity factor β = v/c.

---

1 [11]Radiation weighing factors, ENS publication, https://www.euronuclear.org/info/encyclopedia/r/radiation-weightfactor.htm

Cosmic γ-rays are emitted mostly from the galactic plane and imaged across the sky as a strip along the Milky Way with their maximum intensity in the direction to the center of our galaxy [17]. Some local bright sources such as Crab nebula can add to the γ-rays intensity. Intensity of galactic γ-rays exponentially decreases in the energy range between 10 to1000 MeV. The spectrally integrating flux of cosmic γ-photons is about 10 $m^{-2}$ $s^{-1}$ $sr^{-1}$. Most γ-rays are absorbed by the Earth's atmosphere except for the most energetic quanta. For the rocket velocity below 0.7*c*, the flux of γ-rays will not differ significantly from the flux incident on Earth atmosphere and the radiation danger from galactic γ-rays seems to be not a big concern unless a starship gets close to a local source of intense γ-radiation.

*3.3 Radiation impact on electronic components*

Every high-energy nucleon passing through an electronic component inevitably produces free electrons, i.e. it deposits some electric charge in the semiconductor material producing parasitic signals and causing bits to flip, latch up, or burn out in computer memory. This deposition of charge can "upset" the memory circuits, and the upset rate of a particular part of electronic equipment caused by cosmic radiation in the vicinity of Earth can vary from 10 per day for commercial RAMs to 1 every 2800 years for radiation-hardened RAMs (radiation-hardened component is a device specially designed to resist nucleonic radiation). Two most dangerous effects can cause degradation of electronics: a) Total Dose Effect which is the change of electrical properties of components upon their prolonged exposure to radiation and b) Displacement Damage which occurs when the nucleons slow down and nearly come to rest at the end of their penetration depth, where they deposit the bulk of their kinetic energy and knock semiconductor atoms out of their proper locations in crystal lattice creating defects in a crystal structure capable of trapping the conduction electrons. The laboratory tests of electronic components irradiated by protons and heavy ions were performed by LaBel et al. [18, 19]. SEEs (single event effects) and other effects were detected virtually in all devices bombarded by heavy ions and some showed SEEs under proton irradiation. The cumulative effects such as degradation of current transfer ratio, reference voltage degradation, functional failure, and displacement damage were commonly observed under proton fluences above $10^{11}$ cm-$^{2}$ protons.

The headwind of hydrogen atoms at a rocket speed above 0.3c in the local low-density cavity exceeds $3\times10^{9}$ $cm^{2}$ $s^{-1}$ therefore the unshielded electronic components will degrade to an inoperable condition in minutes of exposure. Hence, a frontal shield against nucleonic radiation of the oncoming relativistic headwind is equally necessary for unmanned (robotic) and manned relativistic spacecrafts. Any relativistic spacecraft or space module, no matter how small or big, must be shielded from the oncoming high-energy nucleons. Cosmic rays seem to be not of great concern for radiation-hardened electronics regarding SEEs during relatively short missions of years in flight but the effect of cumulative degradation of electronic components can be a significant damaging factor in long-range flights of tens of years and more without a proper protection shield.

3.4 *Interstellar dust*

Concentration of interstellar dust grains with their sizes from $10^{-5}$ to $10^{-6}$ m (1 to 10 μm) and their mass from $10^{-17}$ to $10^{-20}$ kg is about $10^{-8}$ $m^{-3}$ in the local low-density cavity [5]. It can increase by the factor of thousand or more in the dense clouds of the galactic arms. The oncoming dust will bombard the frontal parts of the rocket with a rate from 1 to 10 $m^{-2}$ $s^{-1}$, if $\beta > 0.3$. Despite their smallness, the grains can pierce through the frontal protective shield damaging the magnetic coils, walls, and frontal parts of the rocket, making micro-holes in the worst scenario or sputtering the shield material and the rocket hull. The impact of the relativistic multi-atomic grains on the materials has never been studied because we do not possess a means for accelerating the multi-atomic granules to relativistic velocities.

To what type of hazard we can relate the oncoming flux of relativistic dust granules by their influence on materials, electronics, and tissue is not clear. Should we consider them as solid projectiles depositing their kinetic energy into materials and producing a mechanical damage like riffle bullets? Or maybe we have to treat them as the lumps of densely packaged nuclei and electrons ionizing and displacing the atoms and molecules in the target like nucleonic radiation? Referring to our experience with the common kinetic projectiles such as small-shots, bullets, cannon shells, etc. we are inclined to consider the relativistic granules as producing some mechanical damage to materials and tissues. At a relativistic speed however, kinetic energy of each atom in the grain significantly exceeds potential energy of interatomic ties in the lattices of all known materials. The atomic ties of electrons with nuclei in the dust grain and the rocket hull should by be disrupted in their collision. Apparently, each relativistic dust grain with its kinetic energy of hundreds MeV per atom can be better treated as a micro-drop of plasma consisting of nuclei and electrons incident on another dense plasma also consisting of nuclei and electrons. In this case, a portion of atomic electrons will be stripped away from the dust granules in their passage through the frontal shield (electron stripper). The granule becomes an electrically charged micro-drop of plasma and we can hope on its deflection away of the rocket body by the magnetic field of the frontal magnetic shield.

May be, the nuclei of a grain will scatter on the nuclei of the shield in agreement with the relativistic Coulomb scattering. There is no theory of relativistic grain collision with material targets and it is not clear if we can effectively protect a relativistic rocket against the oncoming flow of relativistic dust without a thick and massive dust-wearing bulge made of a solid material in front of the rocket. Possibly, a relatively thin shell of constantly renewable material such as a layer of freezing ice permanently grown on a mesh of thin tubes with refrigerating liquid can compensate the loss of material due to sputtering by the dust granules while serving an electron stripper for the neutral atoms in the oncoming relativistic gas. Obviously, the frontal shield would be the most vulnerable part of relativistic spacecraft.

In addition to gas and dust, interstellar space contains multi-atomic molecules such as polycyclic aromatic hydrocarbons and even fullerens [20] that fill the gap between gas and dust. Every galaxy including our Milky Way is filled with a dusty gas as the necessary component directly participating in star formation and evolution of the galaxies. Regardless of the mode of thrust production and mass of interstellar module, no relativistic flight can be undertaken without a proper protection of crew (if any), electronics, and construction elements against the oncoming relativistic flow of the components of interstellar medium.

## 4. Radiation hazard during the rocket braking

There is another circumstance regarding the relativistic interstellar flights commonly omitted in publications namely rocket protection against relativistic flow of interstellar gas and dust during rocket braking. The frontal shield is able to perform its protective duty during acceleration and following cruising with a constant relativistic speed, i.e. when the rocket nose is directed forward and the protective shield is positioned in front of the rocket. Inevitably, the moment will come to start braking and to cancel the rocket speed upon arrival to destination. In order to start braking, the rocket must be either turned around as a whole by 180 degrees or rearrange its part as a toy transformer to redirect the efflux jet strictly forward. Since protective shield cannot now be placed now in front of the rocket and obscure the exhaust jet, there are two options: a) we risk to turn the whole rocket by 180 degrees exposing its whole rocket body to the full fury of oncoming relativistic interstellar gas and dust without the protective shade of a frontal shield or b) the rocket is transformed so that the vulnerable parts (crew quarters, control rooms, radiators, etc.) are still kept in the shade of the frontal shield while the propulsion beams of ions are redirected forward. The first maneuver would leave the propulsion engine and other parts of the rocket without protection against the relativistic headwind of gas and dust unless the forward-directed efflux jet is capable of sweeping away all atoms, ions, and dust granules in front of the rocket.

At a relativistic velocity, no gas dynamics is applicable to estimate the ion jet sweeping ability. To evaluate the action of the jet ions on interstellar gas molecules, atoms, and ions, the processes of atomic ionization and Coulomb scattering [1, pp. 113–116] must be considered. Hence the efflux jet of high-energy ions emitted from accelerators is to be neutralized by electrons to avoid charge accumulation on the rocket body, the forward-directed efflux is actually a relativistic jet of plasma piercing through the interstellar gas with its map-velocity $\beta_{jet} = (\beta + \beta_i)/(1 + \beta\beta_i)$ according to the relativistic addition equation, where $\beta$ is the relativistic factor of the rocket map-velocity and $\beta_i$ is the proper relativistic velocity factor of the efflux jet of ions and electrons in the rocket coordinate frame. The estimations performed in [1, pp. 114–115] for 1 TW and 100 TW ion propulsion power showed inability of the ion efflux beam to completely ionize the neutral component of interstellar gas and to sweep the ionized interstellar atoms out of the way at the rocket speed above 0.2*c*. The situation with sweeping the dust granules out of the way is even more hopeless so that the propulsion thruster will be under permanent damaging bombardment by relativistic nucleons and duat/ The only possibility to keep all the parts of relativistic rocket in the braking stage behind the protective shield is transformable ion thruster consisting of several ion accelerator units installed symmetrically around the rocket aft, so that each unit can be turned around to redirect its efflux jet almost ahead of the rocket at a small angle with respect to the rocket velocity vector (its axis of symmetry) to avoid any possible damaging affect on the rocket construction elements including the protective shield [5]. This way, the thrust engine still remains in the shade of the protective frontal shield together with the whole rocket while operating in normal regime (Fig. 10). The angle of propulsion jet will be widened a bit in comparison with the acceleration mode resulting in some reduction of thrust but it can be acceptable accounting for a reduced total mass of the rocket by this moment.

It should be mentioned that the shielding system for protection of relativistic rocket or any other relativistic spacecraft cruising in the local low-density cavity can be insufficient in the galactic clouds of much higher density. When we get ready to send an interstellar ship or a module beyond the local cavity, the navigation charts and a map of dence interstellar clouds will be needed for laying out a safe course through the low-density tunnels in the galactic arms.

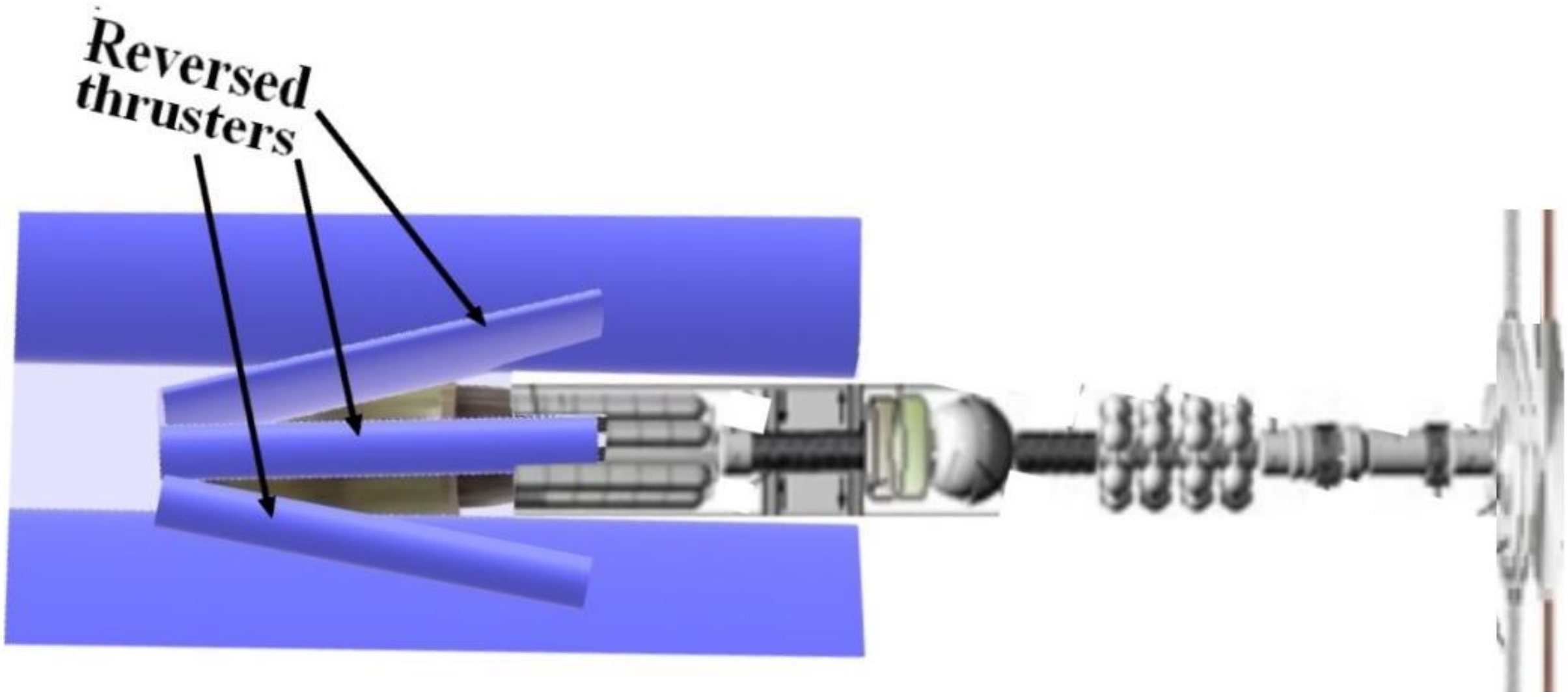

Fig. 10 Conceptual relativistic ion-propulsion rocket (Fig. 8) in braking stage.

**5. Conclusion**

Among the factors potentially limiting our pursuit for unrestrained expansion into the universe, ionizing radiation originated from propulsion engine as well as arising from the very fact of rocket movement with a relativistic speed through interstellar space filled with rarefied gas will be of our highest concern. Despite the extremely low on concentration of gas and plasma in interstellar space, three nucleonic components are hazardous for crew and electronics on board of a relativistic rocket: neutral and ionized components of interstellar gas, cosmic rays, and galactic $\gamma$-radiation. At relativistic rocket velocity, interstellar gas turns into an extremely intense flow of nucleonic radiation incident on the rocket frontal parts. Even at a moderate relativistic speed of $0.2 - 0.4c$, the radiation dose rate absorbed in an unprotected astronaut can be extremely high from hundreds to thousands rems per second. To protect crew and electronics, the proper windward shielding becomes a necessity. Unshielded electronic components would also degrade in minutes of flight at relativistic rocket velocity thus even an unmanned relativistic rocket or a module of any kind will require protection against the nucleonic radiation of oncoming relativistic "headwind". A thick and heavy material shield in front of a spacecraft will hardly be acceptable because of a significant increase in its dry mass. The presence of neutral component in interstellar gas excludes the use of magnetic shielding alone. A combination of electron stripper and magnetic shield seems to be a solution for protection of crew and electronics against radiation produced by relativistic flow of interstellar gas.

Isotropic cosmic rays can be subjected to frontal relativistic beaming in the rocket's coordinate frame, if the rocket moves with a relativistic speed close to the speed of light, so the frontal magnetic shield would absorb or deflect them away partially. At a moderate rocket speed below 0.7c, the relativistic beaming is significantly weaker and not sufficient to significantly reduce the intensity of cosmic rays from the sides and from the aft of the rocket thus an adequate shielding of crew quarters from isotropic cosmic rays will also be needed during the long-term interstellar flights. Owing to its low intensity, cosmic γ-radiation seems to be not of big concern regarding radiation hazard on board of interstellar spacecrafts, at least at moderate rocket velocity. The types of hazardous ionizing radiation and potential radiation sources are listed in Table 1.

**TABLE 1** Most relevant potentially hazardous radiation in relativistic flight. Energies of γ-photons and kinetic energies of massive particles emitted in proton-antiproton annihilation are taken near the maxima of their energy distribution [4]. Kinetic energy of oncoming headwind nucleons $E_k = mc^2(\gamma - 1)$ is a function of rocket velocity v through the factor $\gamma = 1/(1-v^2/c^2)^{1/2}$.

| Radiation origin | Radiated particles | Particle energy (MeV) |
|---|---|---|
| Rocket propulsion engine: | | |
| Photon rocket | γ-photons | 0.511 |
| Meson rocket | γ-photons | 200 |
| | π-mesons | 250 |
| | μ-mesons | 190 |
| Annihilation reactor | γ-photons | 200 |
| | π-mesons | 250 |
| | μ-mesons | 190 |
| Relativistic headwind of gas at v > 0.3c | Electrons | >0.025 |
| | H ions (protons) | >50 |
| | Helium ions (alpha particles) | >200 |
| Cosmic rays | Protons and alpha mostly | 100 – 1000 |
| Galactic γ-radiation | γ-photons | 10 – 1000 |

In addition to nucleonic and γ-radiation, interstellar dust can cause the mechanical damage of the frontal parts of a relativistic spacecraft. The protective shield against nucleonic radiation of interstellar gas headwind will be the most vulnerable to dust bombardment. At relativistic speed, dust granules can be rather considered as the dense lumps of plasma of high-energy nucleons and electrons, which collide with nuclei of the frontal shield or rocket hull knocking atoms from their position in the lattice and producing some secondary mesonic radiation. The radiation hazard from oncoming relativistic headwind and the damaging sputtering of the rocket elements by relativistic interstellar dust are among the most serious problems to be solved before attempting a relativistic flight to the stars.